\documentclass[twocolumn,letterpaper]{revtex4-1}

\usepackage[T1]{fontenc}       

\usepackage{xcolor}
\usepackage{textcomp}

\usepackage[normalem]{ulem}
\usepackage{amsmath}
\usepackage{amssymb}
\usepackage{graphicx}

\newcommand*{\gD}{\Delta}

\newcommand*{\pt}[1]{\left( #1\right)}

\newcommand*{\red}[1]{\textcolor{red}{#1}}

\begin{document}

\title{How did prebiotic polymers become informational foldamers?}

\author{Elizaveta A Guseva}
 \email{eguseva@tutanota.com}
\affiliation{Laufer Center for Physical and Quantitative Biology, \& 
Departments of Physics and   Astronomy and Chemistry, Stony Brook University, Stony Brook, NY, 
(United States)}

\author{Ronald N Zuckermann}
 \affiliation{Lawrence Berkeley National Laboratory (LBNL), Berkeley, CA (United States)}
\author{Ken A Dill}
\email{dill@laufercenter.org}
\affiliation{Laufer Center for Physical and Quantitative Biology, and Departments of Physics \& 
 Astronomy and Chemistry, Stony Brook University, Stony Brook, NY, (United States)}


\begin{abstract}
A mystery about the origins of life is which molecular structures -- and what spontaneous 
processes -- drove the autocatalytic transition from simple chemistry to biology?  Using the HP 
lattice model of polymer sequence spaces leads to the prediction that random sequences of 
hydrophobic ($H$) and 
polar ($P$) monomers can collapse into relatively compact structures, exposing hydrophobic 
surfaces, acting as primitive versions of today's protein catalysts, elongating other such HP 
polymers, as ribosomes would now do. Such foldamer-catalysts form an autocatalytic set, growing 
short chains into longer chains that have particular sequences. The system has capacity for the 
multimodality: ability to settle at multiple distinct 
quasi-stable states characterized by different groups of dominating polymers. This is a testable 
mechanism that we believe is relevant to the early origins of life.
\end{abstract}

\maketitle

\section*{Introduction} 

 Among the most interesting and mysterious processes in chemistry is how the spontaneous 
transition 
occured, more than 3 billion years ago, from a soup of prebiotic molecules to living systems.  
What 
was the mechanism of the Chemistry-to-Biology (CTB) transition?  This is one of those scientific 
puzzles for which theory may be required to preceed experiments, even to suggest what mechanisms 
might be plausible.  In this paper we develop a model to 
explore what prebiotic polymerization processes might have produced long chains of protein like or 
nucleic acid like molecules~\cite{Joyce1987,Abel2005}.  What polymerization processes are 
autocatalytic?  How could they have produced chains that are longer than are currently observed in 
model prebiotic experiments?  And, how might random chain sequences have become informational and 
self-serving?  Our questions here are about physical spontaneous mechanisms, not about the 
different chemistries of monomers or polymer types or polymerization conditions, \emph{per se}.  
 
 \section*{The Chemistry-to-Biology transition has been postulated to entail an autocatalytic 
process}
  Early on, it was 
recognized 
that the transition to biological self-supporting behavior requires autocatalysis, \emph{i.e.} some 
form of positive feedback or 
bootstrapping in which some molecules become amplified and self-sustaining relative to other 
molecules 
~\cite{eigen1971selforganization,Eigen1977,Eigen1978,Dyson1985,Prigogine1989,Kauffman1986}.  That 
work has led to the idea of an \emph{autocatalytic set}, a collection of entities in which any one 
entity can be autocatalytic for another.  We review here some of the key subsequent results, first 
from theory and modeling.  A class of models called GARD (Graded Autocatalysis Replication 
Domain)~\cite{segre1998graded,Segre2000,Markovitch2012} predicts that artificial autocatalytic 
chemico-kinetic networks can lead to self-replication, with a corresponding amplification of some 
chemicals over others. Such systems  display some degree of 
inheritance and adaptability.  GARD models are a subset of `metabolism-first' 
mechanisms, which envision that 
small-molecule chemical processes precede information transfer and precede the first biopolymers.  
Focusing on polymers, Wu and Higgs~\cite{Wu2009} developed a model of RNA chain-length 
autocatalysis.  They envision that some of the RNA chains can spontaneously serve as polymerase 
ribozymes, leading to autocatalytic elongation of other RNAs.  A related model asserts that 
autocatalytic chain elongation arises from template-assisted ligation and random 
breakage~\cite{Tkachenko2014}.  These are models of the `pre-informational' world before 
heteropolymers begin to encode biological sequence-structure relationships.  
 
 Another class of models describes a `post-informational' heteropolymer world, in which there is 
already some tendency of chains to evolve.  In one such model, it is assumed that polymers serve 
as 
their own templates because of the ability of certain heteropolymers to concentrate their own 
precursors~\cite{nowak2008prevolutionary,Ohtsuki2009,Chen2012,Derr2012}.  It supposes an ability 
of 
molecules to recognize ``self'', although without specifying exactly how.  In another such 
model~\cite{Walker2012}, chains undergo sequence-independent template-directed replication.  It 
indicates that functional sequences can arise from non-functional ones through effective 
exploration of sequence space.  These post-informational models predict that template-directed 
replication will enhance sequence diversity~\cite{Derr2012}.   These are abstract models that 
address matters of principle, rather than questions of what particular molecular structures might 
explain the autocatalytic step. Nor do they address the heteropolymeric or informational aspects 
of 
the chains.
 
 There has also been much experimental work, leading, for example, to the creation of artificial 
autocatalytic sets in the laboratory~\cite{VonKiedrowski1986,Lincoln2009,Vaidya2012}. Such systems 
are designed so that pairs of molecules can catalyze each other (i.e. autocatalysis), leading to 
exponential growth of the autocatalytic members.  For example, mixtures of RNA fragments are shown 
to self-assemble spontaneously into self-replicating ribozymes that can form catalytic networks 
that can compete with others\cite{Robertson2014}.  One limitation, however, is that these are 
fragments 
taken from existing ribozymes, so they don't explain the origins from more primitive and random 
beginnings.
 
  Here, we describe a theoretical model that seeks to bridge from the pre- to post-informational 
world, across the Chemistry-to-Biology transition.  We seek a physical basis for how short chains 
could have led to longer chains, for how random chains led to specific sequences, and for a 
structural basis and plausible kinetics for a prebiotic autocatalytic transition.
   
 \section*{The `Flory Length Problem': polymerization processes produce mostly short chains}
 \label{sec:flory} 

Prebiotic polymerization experiments rarely produce long chains.  It is commonly assumed that the 
chain lengths of proteins or nucleic acids that could have initiated the transition to biology must 
be at least 30-60 monomers long~\cite{Szostak1993}.  
Both amino acids or nucleotides can polymerize under prebiotic conditions without enzymes, but 
they 
produce mostly short chains~\cite{Shock1992,Martin1998,PAECHT-HOROWITZ1970,Leman2004a,Orgel2004}.  
Leman et al. showed that carbonyl sulfide (COS), a simple volcanic gas, brings about the formation 
of oligo-peptides from amino acids under mild conditions in aqueous solution in minutes to hours. 
But the products are mainly dimers and trimers~\cite{Leman2004a}.  Longer chains can sometimes 
result through adsorption to clays~\cite{Rao1980,Lambert2008} or 
minerals~\cite{Bernal1949,Ferris1996}, from evaporation from tidal pools~\cite{Nelson2001}, from 
concentration in ice through eutectic melts~\cite{Kanavarioti2001}, or from 
freezing~\cite{Bada2004} or temperature cycling.  Even so, the chain-length extensions are modest. 
 
For example, mixtures of Gly and Gly$_2$ grow to about 6-mers after 14 
days~\cite{Rode1997,Rode1999} on mineral catalysts such as calcium montmorillonite, hectorite, 
silica or alumina.  Or, in the 
experiments of Kanavarioti, polymers of oligouridylates are found up to lengths of 11 bases long, 
with an average length of 4 \cite{Kanavarioti2001} after samples of phosphoimidazolide-activated 
uridine we frozen in the presence of metal ions in dilute solutions.  Similar results are found in 
other polymers: a prebiotically plausible mechanism produces oligomers having a combination of 
ester and amide bonds up to length 14~\cite{Forsythe2015}.  

\begin{figure}[ht!]
  \centering
  \includegraphics[width=0.9\columnwidth]{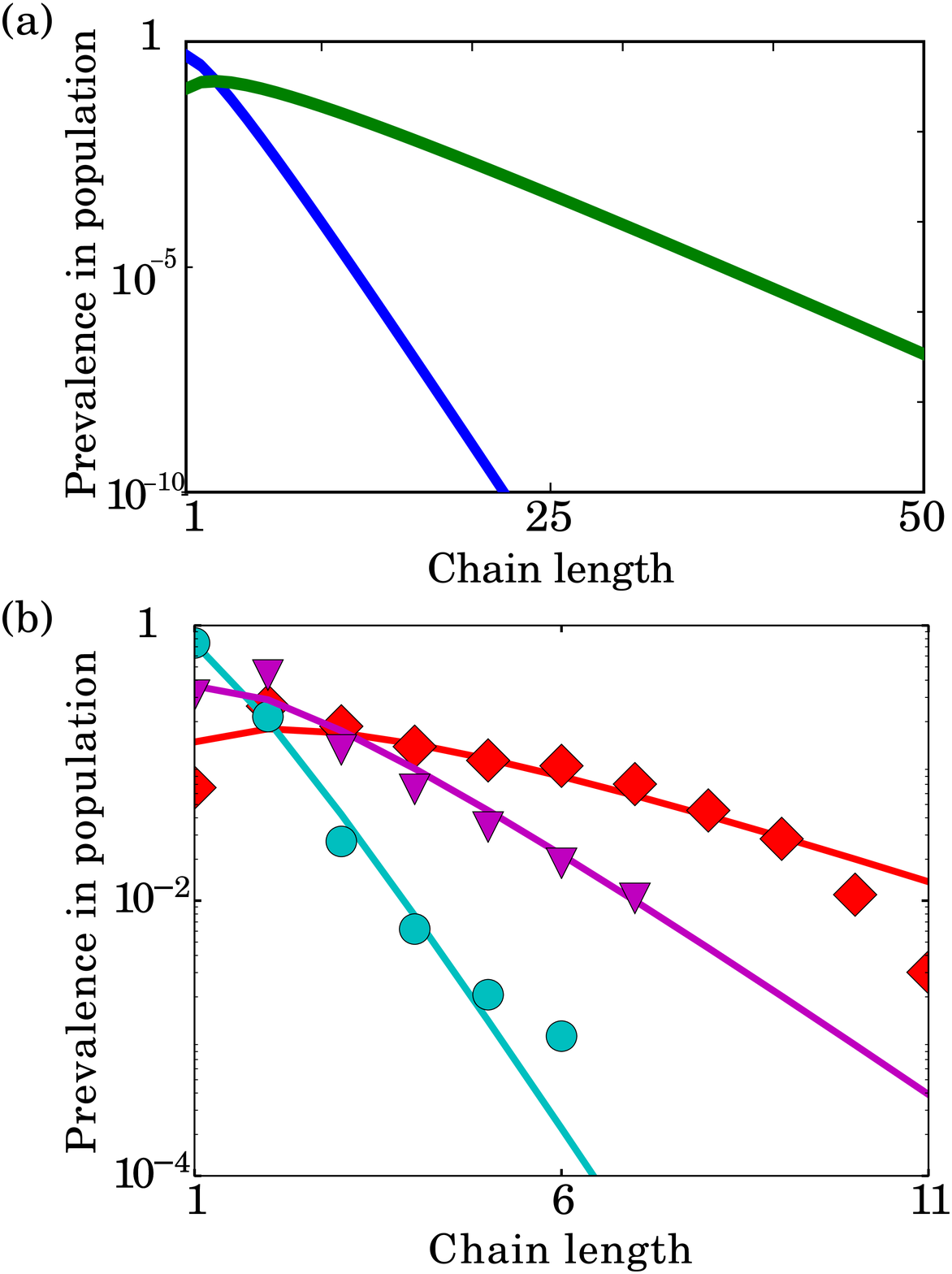} 
  \caption{\textbf{Polymerization processes lead to mostly short chains.} (a)  Spontaneous 
polymerization processes typically lead to a Flory distribution of chain lengths. 
Green line gives $\langle  l \rangle= 6$, blue corresponds to $\langle l \rangle=2$
(b) Fitted distributions from experiments on prebiotic polymerization: red -- 
Kanavarioti~\cite{Kanavarioti2001}, cyan -- Ding~\cite{Ding1996}, 
magenta -- Ferris~\cite{Ferris1999}}
  \label{fig:flory}
\end{figure}

It is puzzling how prebiotic processes might have overcome what we call the ``Flory Length 
Problem'' -- i.e. the tendency of any 
polymerization process to produce a distribution in which there are more short chains and fewer 
long chains.  Standard polymerization mechanisms lead to the the Flory or Flory-Schulz distribution 
of populations $f(l)$, whereby short chains are exponentially more populated than longer 
chains~\cite{Flory1953}, 
\begin{equation}
 f(l)=a^2l(1-a)^{l-1},\label{eq:flory}
\end{equation} 
where $l$ is the chain length and $a$ is the probability that any monomer addition is a chain 
termination.  The average chain length is given by $\langle l \rangle = a(2- a)$; see Figure 
\ref{fig:flory}(a).  

 Prebiotic monomer concentrations are thought to have been in the range of micromolar to 
millimolar~\cite{Stribling1987,Huber1998,Aubrey2009,Kanavarioti2001,Lazcano1996}.  Given micromolar 
concentrations of monomers, and given $\langle l \rangle = 2$, the concentration of 40-mers would 
be $\approx 10^{-19} $ mol/L.  Figure~\ref{fig:flory}(b) shows that where the chain-length 
distributions are known for prebiotic syntheses, they are well fit by the Flory 
distribution (or exponential law $f(l)\propto$ 
constant$^l$)~\cite{nowak2008prevolutionary,Derr2012}).

\section*{The foldamer-autocat mechanism: Short HP chains fold and catalyze the elongation of 
other 
HP chains}

 We propose that the key to the Chemistry-to-Biology transition may have been \emph{foldable 
polymers} (`foldamers').  Today's biological foldamers are predominantly proteins (although RNA 
molecules 
and synthetic polymers can also fold~\cite{Gellman1998,Lee2005a,Capriotti2008}).  Many foldamers 
adopt specific native conformations, mainly through a binary solvation code of particular sequence 
patternings of the $H$ (hydrophobic) and $P$ (polar) monomers~\cite{Chan1991}.  We call these $HP$ 
copolymers.  

 Since today's bio-catalysts are proteins, it is not hard to imagine that yesterday's primitive 
proteins could have been primitive catalysts.  Precision and complexity are not required for 
peptides to perform biological functions. For example, proteins generated from random libraries 
can 
sustain the growth of living cells~\cite{Fisher2011}.  And, specific binding actions between 
random 
peptides and small molecules are not rare~\cite{Cherny2012}.  Below, we describe results of 
computer simulations that lead to 
the conclusion that short random HP chains carry within them the capacity to autocatalytically 
become longer and more protein-like.  

\subsection*{Here are the premises of the model}
 
 \begin{enumerate}
 \item Some random HP sequences can fold into compact structures.
 \item Some of those foldamers will have exposed hydrophobic ``landing pad'' surfaces.
 \item Foldamers with landing pads can catalyze the elongation of other HP chains.
 \item These foldamer-catalysts form an autocatalytic set.
\end{enumerate}
Here is evidence for these premises.
\begin{enumerate}
\item Non-designed random $HP$ sequences are known to fold.  $HP$ polymers have been studied 
extensively as a model for the folding and evolution of 
proteins~\cite{lau1989lattice,Chan1991,Miller1995,Yue1995,agarwala1997local}.  Those studies show 
that folded structures can be encoded simply in the binary patterning of polar and hydrophobic 
residues, with finer tuning by specific interresidue contacts~\cite{Yue1992,Xiong1995}.  This is 
confirmed by experiments~\cite{Lim1991,Kamtekar1993,Wei2003,Brisendine2015}. 

\item Exposed hydrophobic clusters and patches are common on today's proteins.  A study of 112 
soluble monomeric proteins~\cite{Lijnzaad1996} found patches ranging from $200$ to 
$1,200$\textit{\AA}$^2$, averaging around $400$\textit{\AA}$^2$; they are often binding sites for 
ligands or other proteins.  Modern proteins have many sites of interaction with other proteins, 
typically nearly a dozen partners.  Almost 3/4 of protein surfaces have 
geometrical properties that are amenable to interactions and those sites are enriched in 
hydrophobes~\cite{Tonddast-Navaei2015}.

\item  Surface hydrophobic patches on proteins are often sites of 
catalysis~\cite{MitchellGuss1983,Lijnzaad1996,VanEe1997,Witt1998}. For example, hydrophobic 
clusters on the surface of lipases serve as initiation sites where the hydrophobic tail of a 
surfactant interacts with the patch first~\cite{VanEe1997}. A hydrophobic cluster on Cytochrome-c 
Oxidase is known to increase $k_{cat}$\cite{Witt1998}. 

\item  Primitive proteins might have catalyzed peptide-chain elongation.  Of course, today's cells 
synthesize proteins using ribosomes, wherein the catalysis is carried out by RNA molecules.  Yet, 
there are reasons to believe that peptide chain elongation might alternatively be catalyzable by 
proteins.  First, peptide chain elongation entails a condensation step and the removal of a water 
molecule~\cite[chapter 3, p.~82]{Nelson2008}.  Dehydration reactions can occur in water if carried 
out in nonpolar environments~\cite{Manabe2001,Manabe2002}, such as protein surfaces.  Second, a 
major route of protein synthesis in simple organisms such as bacteria and fungi utilizes 
nonribosomal peptide synthetases, and which don't involve 
mRNAs~\cite{Stachelhaus1998,marahiel2009working}.
\end{enumerate}

\section*{Modeling the dynamics of HP chain growth and selection}
 
 \textbf{The dynamics of the model.}  We assume that chain polymerization takes place within a 
surrounding solution that contains a sufficient supply of activated $H$ and $P$ monomers.  Since 
living systems -- past or present -- must be out-of-equilibrium, this 
assumption is not very restrictive.  In our model, activated $H$ and $P$ monomers are supplied by 
an external source at rate $a$.  A given chain elongates by adding a monomer at rate $\beta$.  
Just 
to keep the bookkeeping simple, we consider a steady state process in which molecules are removed 
from the system by degradation or dilution at the same rate they are 
synthesized.  We assume chains can undergo spontaneous hydrolysis due to interaction with water; 
any bond can be broken at a rate $h$. Without loss of generality we define the unit rate by 
setting 
$\beta = 1$.  All other rates are taken relative to this chain-growth rate.
 
 \begin{figure}[ht!]
  \centering
  \includegraphics[width=0.8\columnwidth]{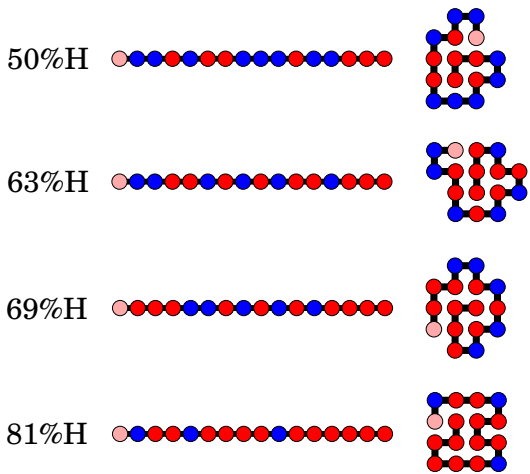} 
  \caption{\footnotesize{\bf{Examples of HP sequences that fold to unique native structures in the 
HP lattice model.} Red (or pink if in the beginning of the sequence) corresponds $H$ monomers and, 
blue to $P$.}}
  \label{fig:hydro-effect}
\end{figure}
 \textbf{Chain folding in the model.}  In addition, our model also allows for how the collapse 
properties of the different HP sequences affect the populations that polymerization produces.  A 
standard way to study the properties of HP sequence spaces is using the 2D HP lattice 
model~\cite{lau1989lattice,Chan1991}.  In this model, each monomer of the chain is represented as 
a 
bead.  Each bead is either H or P.  Chains have different conformations, represented on a 
2-dimensional square lattice.  The free energy of a given chain in a given conformation equals 
(the 
number of HH noncovalent contacts) $\times$ (the energy $e_H$ of one HH interaction).  Some HP 
sequences have a single lowest-free-energy structure, which we call \emph{native}, having native 
energy $E_{nat}$:

\begin{equation}
 E_{nat}=n_{h\phi}e_H.
\end{equation} 

 where $n_{h\phi}$ is the number of HH contacts in the native structure of that particular 
sequence.  
 
 A virtue of the HP lattice model is that for chains shorter than about 25 monomers long, every 
possible conformation of every possible sequence can be studied by exhaustive computer 
enumeration. 
 Thus folding and collapse properties of whole sequence spaces can be studied without bias or 
parameters.  Prior work shows that the HP lattice model reproduces many of the key observations of 
protein sequences, folding equilibria, and folding kinetics of proteins\cite{Dill1999}.  A main 
conclusion from previous studies is that a non-negligible fraction of all possible HP sequences 
can 
collapse into compact and 
structured and partially folded structures resembling native proteins~\cite{lau1989lattice}; see 
fig.~\ref{fig:hydro-effect}.  The reason that the 2-dimensionality adequately reflects properties 
of 3-dimensional proteins is because the determinative physics is in the surface-to-volume ratios 
(because the driving force is burial of H residues).  And, it is helpful that the 10-30-mers that 
can be studied in 2D have the same surface-to-volume ratios as typical 3D proteins, which are 
100-200-mers~\cite{Giugliarelli2000}.

 We assume that folded and unfolded states behave differently, as they do in modern proteins.  We 
suppose that a folded chain is prevented from further growth, and also are protected from 
hydrolysis.  This simply reflects that open chains are much more accessible to degradation from 
the 
solvent or adsorption onto surfaces than are folded chains.  Even so, folding in our model is a 
reversible, as it is for natural proteins, so some small fraction of the time even folded chains 
are unfolded, and in that proportion, our model allows further growth or degradation.  For this 
purpose, we estimate the folding and unfolding rate coefficients for any HP sequence 
as~\cite{Ghosh2009}:
\begin{equation}
 \ln\pt{\frac{k_f}{k_u}}=-\gD G/kT = E_{nat}/kT-N\ln z,
\end{equation} 
 where $z$ is the number of rotational degrees of freedom per peptide bond.

\begin{figure}[ht!]
  \centering
  \includegraphics[width=0.8\columnwidth]{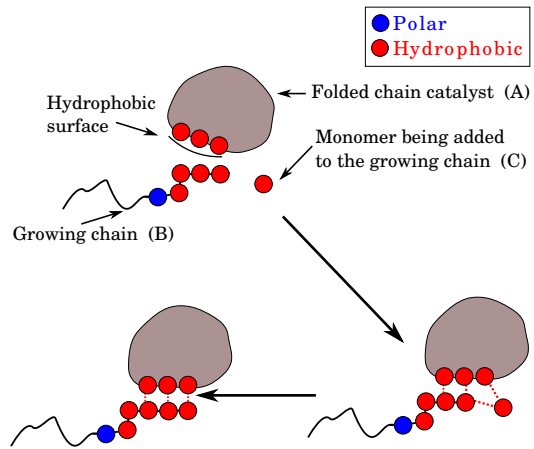} 
  \caption{\footnotesize{\textbf{Some HP foldamers have hydrophobic patches, which serve as 
``landing pads'' that can catalyze the elongation of other HP chains.}  Chain $A$ folds and 
exposes a hydrophobic sticky spot, or landing pad, where another $HP$ molecule $B$, as well as an 
$H$ monomer 
$C$, can bind.  This localization reduces the barrier for adding monomer $C$ to growing chain 
$B$.}}
  \label{fig:hp-catalysis}
\end{figure}

\textbf{Catalysis in the model.}  Some HP sequences will fold to have exposed hydrophobic 
surfaces. 
 These surfaces could act as primitive catalysts, as modern proteins do more optimally today.  
Fig. 
\ref{fig:hp-catalysis} illustrates a common mechanism of catalysts; namely translational 
localization of the reacting components.  A protein $A$ (the catalyst molecule) has a hydrophobic 
`landing pad' to which a growing reactant chain $B$ and a reactant monomer $C$ will bind, 
localizing them long enough to form a bond that grows the chain.  How much rate acceleration could 
such a localization give?  Here is a rough estimate.  

 For chain elongation, the catalytic rate will increase if the polymerization energy barrier is 
reduced by hydrophobic localization, by a factor $\beta_\mathrm{cat}/ \beta_\mathrm{no\,cat} 
\propto \exp(E_{H}\cdot n_{c}/kT)$, where $n_c$ is the number of H monomers in the landing pad 
(see 
figure \ref{fig:fold-cat}).  The free energy of a typical hydrophobic interaction 
is 1-2 $kT$.  We take the minimum size of a landing pad to be 3.  For a landing pad size of 3-4 
hydrophobic monomers, this binding and localization would reduce the kinetic barrier by 3-8 kT, 
increasing the polymerization rate by 10x to 3000x.  Of course, this rate enhancement is much 
smaller than the $10^7$-fold of modern ribosomes~\cite{Sievers2004a}, but even small rate 
accelerations might have been relevant for prebiotic processes.

In order to simulate this dynamics, we run stochastic simulations. 
We used Expandable Partial Propensity Method (EPDM)~\cite{Guseva2016b}. Description and the 
corresponding C++ library, can be found at: 
https://github.com/abernatskiy/epdm. 
  
\section*{Results}
\begin{figure}[ht!]
  \centering
  \includegraphics[width=\columnwidth]{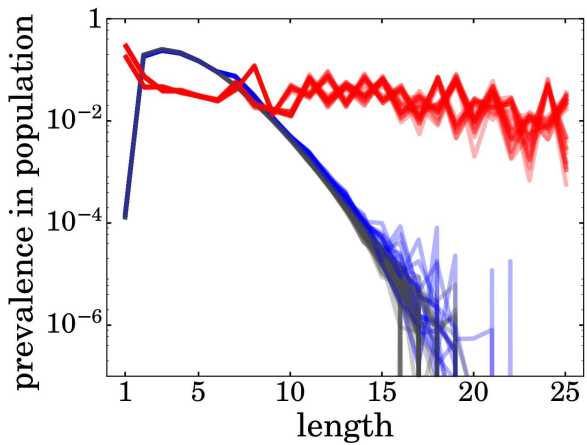}
  \caption{\footnotesize{\bf{Chains become elongated by foldamer-catalyst HP sequences.}  
\textbf{Case 1 (gray):} A soup of chains has a Flory-like length distribution in the absence of 
folding and catalysis.  \textbf{Case 2 (blue):} A soup of chains still has a Flory-like length 
distribution in the absence of catalysis (but allowing now for folding).  \textbf{Case 3 (red):} A 
soup of chains contains considerable populations of longer chains when the soup contains HP chains 
that can fold and catalyze.  We run 30 simulations for every case. To produce each line we took a 
time average over $10^6$ time points in the steady state interval, then counted molecules for each 
length and divided it it by the total molecular count.}}
  \label{fig:sim.flory-fold}
\end{figure}
\subsection*{Folding alone does not solve the Flory Length Problem.  But folding plus catalysis 
does.}
\begin{figure*}[htb!]
  \centering
  \includegraphics[width=\textwidth]{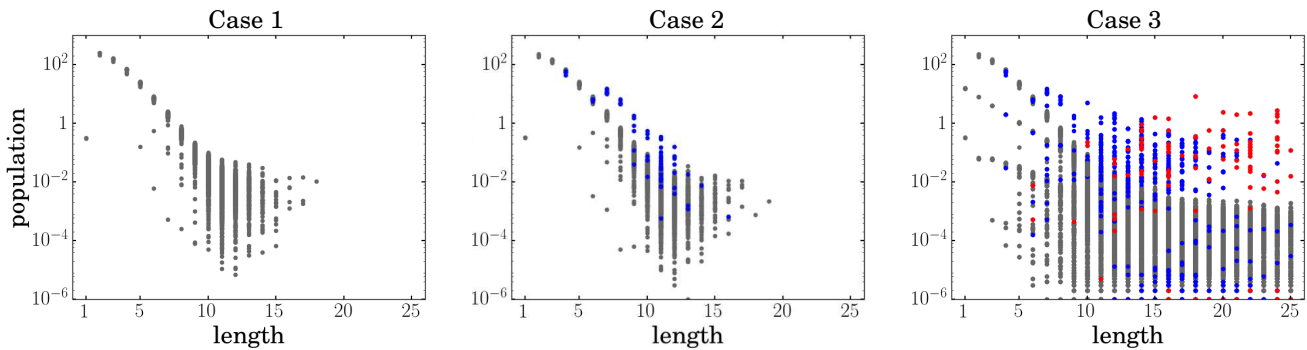}
  \caption{\footnotesize{\textbf{\bf{The distributions over individual sequences are highly 
heterogeneous.}  We show the populations (molecule counts of individual 
sequences) for the three cases: in case 1 we don't allow folding or catalysis, in case 2 
we allow folding but not catalysis, and in case 3 both folding and catalysis are allowed. For 
all the cases gray dots represent populations of the sequences that cannot fold, 
blue -- sequences that fold, but cannot 
catalyze and red -- sequences which act as catalysts and for which at least one elongation 
reaction 
has been catalyzed.  For cases 1 and 2, populations of the sequences of the given length are 
distributed exponentially. Thus we can take mean or median population for the given length as a 
faithful representation of the behavior of average sequence of that length. The case 3 is 
drastically different: the populations of the sequences of the given lengths are distributed 
polynomially. While most of the sequences have very low population for the longer chains, 
several sequences (mostly autocatalytic ones) have very high ones and constitute most of the 
biomass. 
For the case 3 neither mean or median are good representations of the behavior of the chains, 
as we can see from the figure, all the chains basically separate into two groups with different 
distributions, this information cannot be shown in the mean or median.  Every 
point on the panels is a time average over $10^6$ time points in the steady state interval. Lower 
limit of $10^{-6}$ is due to computational precision.}}}
  \label{fig:stats-scatter-018}
\end{figure*}
 We compare three cases:  Case 1 is a reference test in which sequences grow and undergo 
hydrolysis 
but no other factors 
contribute, Case 2 allows for chain folding, but not for catalysis and Case 3 allows for both 
chain 
folding and catalysis.  Case  1 simply recovers the Flory distribution, as expected, with 
exponentially decaying populations with chain 
length (see figure \ref{fig:sim.flory-fold} gray lines).  In the Case 2 when chains can fold, they 
can bury some monomers in their folded cores.  So, chains that are compact or folded degrade more 
slowly than chains that don't fold.  However, 
shows that this situation does not solve the Flory length problem either.  Folding does increase 
the 
populations of some foldamer sequences relative to others, but the effects are too small affect 
the 
shape of the overall distribution (see figure \ref{fig:sim.flory-fold} blue lines).  
Case 3 gives considerably larger populations of longer chains than cases 1 or 2 give (red lines on 
figure~\ref{fig:sim.flory-fold}).  When chains 
can both fold by themselves and also catalyze the elongation of others, such polymerization 
processes will ``bend'' the Flory distribution.  This effect is robust over an order of 
magnitude of the hydrolysis and dilution parameters.  The result is that some HP chains can fold, 
expose some hydrophobic surface, and reduce the kinetic barrier for elongating other chains.  
These 
enhanced populations of longer chains occur even though the degree of barrier reduction is 
relatively small.

Case 3 is qualitatively different than cases 1 and 2.  Even though cases 1 and 2 have substantial 
variances, they have well-defined mean values that diminish exponentially with chain length.  Case 
3 has much bigger variances, and a polynomial distribution of chain lengths, so neither the mean 
nor median are good representations of the behavior of the chains; see figure 
\ref{fig:stats-scatter-018}(Case 3).

\subsection*{The foldamer-catalyst sequences form an autocatalytic set.}

\begin{figure}[htb!]
  \centering
  \includegraphics[width=\columnwidth]{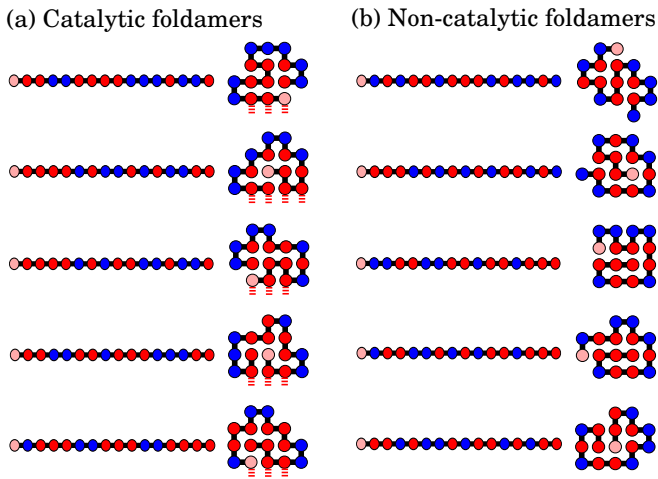} 
  \caption{\footnotesize{\textbf{(a) HP lattice chains that fold and are autocatalytic.}  They 
fold 
into unique structures and have landing pads that can catalyze the elongation of each other.  
\textbf{(b) HP chains that fold, but are not catalytic.}  Most chains are not catalysts, but the 
size of the autocatalytic set is non-negligible; see Fig.~\ref{fig:hp-statistics}.}}
  \label{fig:fold-cat}
\end{figure}

 The present model makes specific predictions about what molecules constitutes the autocatalytic 
set -- which HP sequences and native structures are in it, and which ones are not.  Figure 
\ref{fig:fold-cat} shows a few of the HP sequences that fold to single native structures.  Figure 
\ref{fig:fold-cat} (a) shows those foldamers that are catalysts while Figure \ref{fig:fold-cat} 
(b) 
shows those foldamers that are not catalysts.  

\begin{figure}[ht!]
  \centering
  \includegraphics[width=0.9\columnwidth]{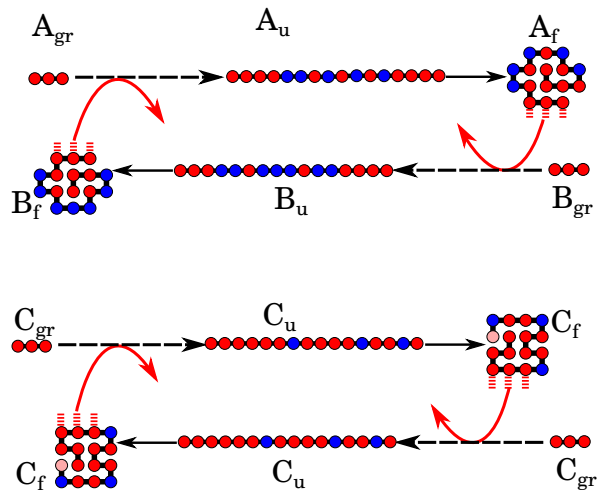}
  \caption{\footnotesize{\bf{Top: Cross-catalysis of 2 different sequences.  Bottom:  
Auto-catalysis of 2 copies of an identical sequence.}  Dashed arrows (\textbf{-\,-\,-}) represent 
multiple 
reactions of chain growth. Among them there are both $\cdots HH+H\to \cdots HHH$ catalyzed 
reactions and spontaneous chain elongations. Catalysis is 
represented by red solid arrows (\red{\textbf{\textemdash}}). Solid black lines 
(\textbf{\textemdash}) are folding reactions. Chains, which we call ``autocatalytic'' experience 
catalysis during one (or more often several) of the steps of elongation. Then, when they reach the 
length at which they can fold ($A_u,\, B_u,\, C_u$), they fold and serve as catalysts them selves 
($A_f,\, B_f,\, C_f$). Mutual catalysis can happen between different sequences (here A and B) and 
between different instances of the same sequence (here C).}}
  \label{fig:kinExamples}
\end{figure}

 In short, all HP sequences that are foldamer-catalysts are members of the autocatalytic set: any 
two HP foldamer-catalyst sequences are autocatalytic for each other.  Figure~\ref{fig:kinExamples} 
shows two examples of autocatalytic paired chain elongations.  The top row of 
Figure~\ref{fig:kinExamples} shows \emph{crosscatalysis}: a polymer $A$ elongates a polymer $B$ 
while $B$ is also able to elongate $A$.  The bottom row of Figure~\ref{fig:kinExamples} shows 
\emph{autocatalysis}: one molecule $C$ elongates a another $C$ molecule in solution.

\subsection*{The size of the autocatalytic set grows with the size of the sequence space.}

\begin{figure}[hbt!]
  \centering
  \includegraphics[width=0.9\columnwidth]{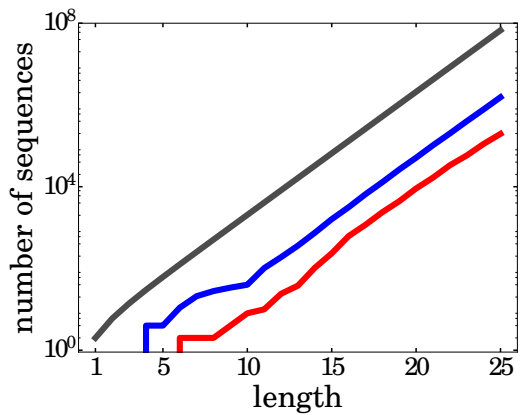} 
  \caption{\footnotesize{\bf{Different sequence spaces grow exponentially with chain length.}  
(Gray) The number of all HP sequences.  (Blue) The number of foldamers.  (Red) The number of 
foldamer catalysts.}}
  \label{fig:hp-statistics}
\end{figure}

\begin{figure}[ht!]
  \centering
  \includegraphics[width=0.9\columnwidth]{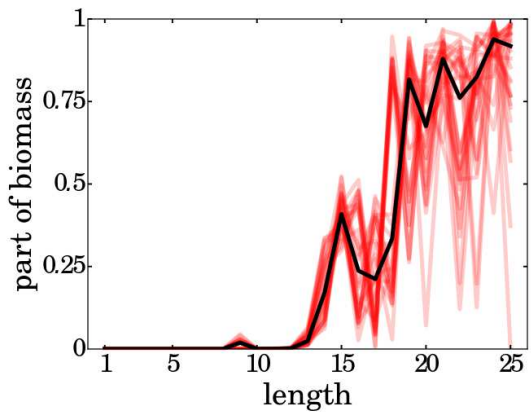} 
  \caption{\footnotesize{\textbf{The longer chains the chains, the bigger the contribution of the 
autocatalysts.} Each red line shows how the contribution of autocatalytic chains to the biomass 
of the given length grows with chain length. Different red lines correspond to different 
simulation runs.  The black line shows the median over 30 simulations.
}}
  \label{fig:biomass}
\end{figure}

 An important question is how the size of an autocatalytic set grows with the size of the sequence 
space.  Imagine first the situation in which the chemistry-to-biology transition required one or 
two `special' proteins as autocatalysts.  This situation is untenable because sequence spaces grow 
exponentially with chain length.  So, those few particular special sequences would wash out as 
biology moves into an increasingly larger sequence space `sea'.  In contrast, 
Figure~\ref{fig:hp-statistics} shows that the present mechanism resolves this problem.  On the one 
hand, the fraction of HP sequences that are foldamers is always fairly small (about 2.3\% of the 
model sequence space), and the fraction of HP sequences that are also catalysts is even smaller 
(about 0.6\% of sequence space).  On the other hand, Figure~\ref{fig:hp-statistics} shows that the 
populations of both foldamers and foldamer-cats grow in proportion to the size of sequence space.  
The implication is that the space of autocats in the CTB might have been huge.
 Figure \ref{fig:biomass} makes a closely related point.  It shows that for longer chains, the 
fraction of biomass that is produced by autocatalysts completely takes over and dominates the 
polymerization process, relative to just the basic polymerization dynamics itself, even though the 
catalytic enhancements are quite modest.  This is due to two factors: (1) the number of 
autocatalysts grow longer sequences (see fig.\ref{fig:hp-statistics} and (2) folding alone is not 
sufficient to populate longer chains.  We find that the average hydrophobicity of the dominant 
sequences in these runs is $68\%$.

\subsection*{Evolvability of HP ensembles}
There are a few problems CTB models in general and autocatalytic models in particular encounter. 
One of them is lack of variability and evolvability. Due to the compositional bias or poor 
dynamical structure of 
the model such systems converge to one state (attractor or attractor basin) determined by 
internal dynamics of the system and do not response to directional 
selection (see discussions in \cite{Derr2012,Vasas2015} for example). For a complex system that 
has many attractors a  perturbation can move the system over a threshold to the basin of another 
attractor. This allows for exploration of the sequence space and thus possible evolvability of the 
system. 

\begin{figure}[ht!]
  \centering
  \includegraphics[width=0.9\columnwidth]{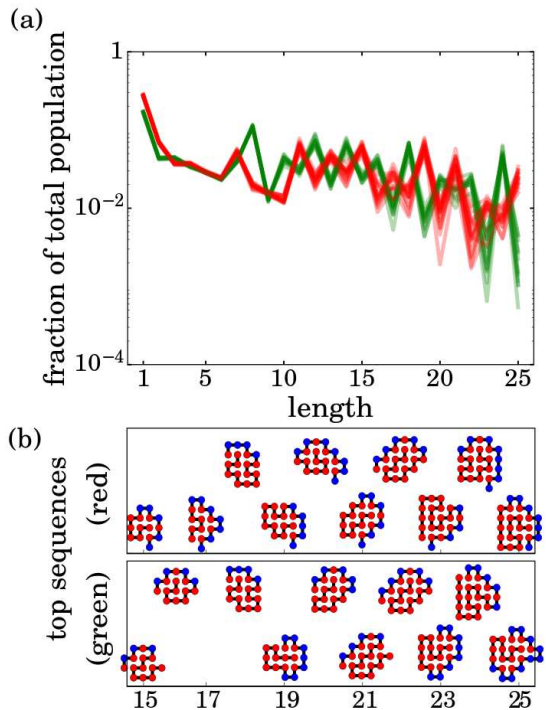}
  \caption{\footnotesize{\textbf{} (a) $HP$ catalytic system has 
at least two attractors. The lines are length distributions from case 3. Again, 
each line represents distribution of length in the steady state for one simulation run. It is 
clear, that there are two kinds of distribution which get realized during the simulations. 
The system 
bifurcates either to a state represented by a green line or to one represented by red one. These 
are the same lines as on figure \ref{fig:stats-scatter-018}(a), but separated in two sets by the 
clustering algorithm k-means. (b) Structure of the sequences which most often
are main contributors into the total population of the polymers of their length. Top panel 
corresponds to the macrostate shown in red on the panel (a), lower one, to the one shown in green. 
}}
  \label{fig:distr1837-dyn}
\end{figure}

HP ensembles have, we believe, two possible attractors, which allow for the exploration 
of the sequence space.  First, as one can see from the figure 
\ref{fig:distr1837-dyn}(a) trajectories split distinctively between two 
attraction distributions. There are no trajectories that lay in between the two attractors, which 
shows that there's no switching between the attractors and the separation is not result of 
stochasticity. In addition to that each of the distribution has a set of specific sequences which 
most often dominate the populations. Figure \ref{fig:distr1837-dyn}(b) shows a few of the 
structures dominating HP ensembles for the ``green'' distribution and for the ``red'' ones.  The 
red and green species differ simply by different starting 
seeds for the simulations.  Each of the two attractors has its own ``signature ensemble'' of HP 
sequences that is an emergent property of the dynamics.  It is possible that adding more realism 
to 
our model (20 monomer types, rather than 2; allowing for longer chains; etc) could lead 
to larger numbers of attractors.
Second, our simulations are limited to 25mers, but in fact the chains can grow longer. This fact 
allows for the further exploration of the sequence and functionality space beyond what can be seen 
in our simulations. If we are talking about protein-like molecules, some of the chains will act 
not only as autocats but also would be capable of binding other molecules, which could result in a 
chemical innovation.

At this point, we note what our model is, and what it is not.  Our model is not intended as an 
 accurate atomistic depiction of a real catalytic mechanism.  It is a coarse-grained toy model, of 
which there will be variants.  The mechanism we explore here is the translational localization of 
the two reactants, polymer $B$ and monomer $C$, in the chain extension reaction.  And, while this 
model is 2-dimensional, extensive previous studies have shown that it captures many important 
principles of folding and sequence-to-structure relationships.  At the present time, this type of 
model is the only unbiased, complete and practical way to explore plausibilities of physical 
hypotheses such as the present one.

We note that the present model is not necessarily exclusive to proteins.  Nucleic acid molecules 
are also able to fold in water, indicating differential solvation.  While our present model 
focuses 
on hydrophobic interactions, it is simply intended as a concrete model of solvation, that could 
more broadly include hydrogen bonding or other interactions.  So, while our analysis here is only 
applicable to foldamers, that does not mean it is limited to proteins.  The unique power that 
foldable molecules have for catalyzing reactions -- in contrast to other nonfoldable polymeric 
structures -- is that foldamers lead to precisely fixing atomic interrelationships in relative 
stable ways over the folding time of the molecule.  It resembles a microscale solid, with the 
capability that substrates and transition states can recognize, bind, and react to those stable 
surfaces. For example, serine proteases utilize a catalytic triad of 3 amino acids.  So, 
foldability 
in some type of prebiotic polymer, could conceivably have had a special role in allowing for 
primitive catalysis.  Here, we use a toy model to capture that simple idea, namely that a folded 
polymer can position a small number of residues in a way that can catalyze a reaction.

\section*{Conclusions}
\label{sec:evolution}
 It has been recognized that life's origins require some form of 
autocatalysis~\cite{Kauffman1986,Dyson1985,Eigen1978}.  But, what molecular structural mechanism 
might explain it?  Here, we find that autocatalysis is inherent in the following process (see 
Figure \ref{fig:kinExamples}):  HP polymers 
are synthesized randomly; a small fraction of those HP polymers fold into relatively stable 
compact 
states; a fraction of those folded structures provide relatively stable `landing pad' hydrophobic 
surfaces; those surfaces can help to catalyze the elongation of other HP molecules having foldable 
sequences.

 The HP model allows for unbiased counting of sequences that do fold, don't fold, or fold and have 
a potentially catalytic hydrophobic landing pad.  A non-negligible fraction of all possible HP 
sequences fold to unique structures ($2.3\% $ for lengths up to 25-mers). The fraction of all 
possible HP sequences that have catalytic surfaces (as defined above) is $12.7\%$ of foldable 
sequences, or $0.3\%$ of the whole sequence space.  These ratios remain relatively constant with 
chain length, at least up to 25-mers; see figure \ref{fig:hp-statistics}.  This and successful 
designs of foldable, biologically active proteins based on the HP folding rule~\cite{Murphy2015} 
suggests that folding in HP polymers is not rare.  
   The present model provides an experimentally testable prediction for what early 
  polymer sequences could be autocatalytic, and provides a structural and kinetic mechanism for 
their action. 

\section*{Acknowledgments}
Authors appreciate support from the Laufer Center as well as from NSF grant MCB1344230. Special 
thanks to Anton Bernatskiy for fruitful discussions.

\newpage
\appendix
 \section*{SI: Simulations}
To perform our stochastic simulations, we first needed to develop appropriate simulation code 
because of the large numbers of different molecular species that must be treated here.  A 
description of the method, called the Expandable Partial Propensity Method (EPDM), and the 
corresponding C++ library, can be found at: 
https://github.com/abernatskiy/epdm~\cite{Guseva2016b}. 
The challenge is to keep track of all the molecular species and to search the full conformational 
spaces of each 
chain.  This is NP-hard.  We use the HP Sandbox algorithm\cite{lau1989lattice,Dill2008a} 
\footnote{A 
Python implementation and description can be found at: 
http://hp-lattice.readthedocs.org/en/latest/}, which is limited to maximum chain lengths of 25 
monomers.  To handle computational limitations, we restricted the total number of species to the 
level of a few thousands.  We impose this limit by introducing a dilution parameter $d$: molecules 
are randomly removed from the system with probability $\propto d$.  Physically, it represents 
molecules that diffuse out of the reaction volume.  The total numbers of molecules within the 
reaction volume vary in the range of $10^2-10^4$.  We start our simulations with a small pool of 
monomers, usually fewer than 100 molecules.  Here are the dynamical steps:

\begin{itemize}
 \item Polymerization happens when monomers react with other monomers or polymers at a rate $\beta 
= 1$:
\begin{equation}
 1\mbox{-mer}+n\mbox{-mer} \xrightarrow{\beta} (n+1)\mbox{mer}
\end{equation}

\item New monomers are imported into the system at high rate $a\gg1$.

\begin{equation}
 \emptyset \xrightarrow{a} H\,\,\mbox{or}\,\,P
\end{equation}

\item We assume a chain can break at any internal site by hydrolysis.  This happens with rate $h$ 
per chain bond. 
\begin{equation}
 n\mbox{-mer} \xrightarrow{h} l\mbox{-mer}+(n-l)\mbox{-mer}
\end{equation}
Typical values for the half-time for the hydrolysis of a bond under neutral conditions and room 
temperature are on the order of hundreds of years\footnote{The hydrolysis rate constants of 
oligopeptides in neutral conditions are of the order of $10^{-11}-10^{-10}$: $1.3  10^{-10} 
M^{-1}s^{-1} $ for benzoylglycylphenylalanine ($t_{1/2} = 128 y$)\cite{Bryant1996}, $6.3  10^{-11} 
M^{-1} s^{-1}$($t_{1/2}=350 y$) for glycylglycine and $9.3 10^{-11}M^{-1} s^{-1}$ for glycylvaline
\cite{Smith1998}.}. Here, we explored a range of hydrolysis rates that are about $0.01-1$ of the 
polymerization rate.  Hence, our model polymerization rates are on the order of days to years.

\item We assume the system becomes diluted, at rate $d$.  This has the practical purpose of 
limiting the total population of polymer in the system.  We explored values of $d$ from $\propto 
0.01- 1\beta$.  Given the values of $a$ we used, it results in $\propto 10^2- 10^4$ chains in the 
simulation volume.  
\begin{equation}
 \mbox{anything} \xrightarrow{d}\emptyset
\end{equation}

\item Folding and unfolding reactions happen much faster than the polymerization processes, with 
corresponding rate coefficients of $k_f\gg k_{u}\gg\beta$:
\begin{equation}
\begin{split}
 \mbox{folded chain}&\xrightarrow{k_u}\mbox{unfolded chain}  \\
 \mbox{unfolded chain}&\xrightarrow{k_f}\mbox{folded chain}
\end{split}
\end{equation}
 We used the most realistic values we could obtain for these rates and for the folding free 
energies for proteins.  We took $E_{nat}$ from the HP model, known folding free energies from 
experimental data~\cite{Ghosh2010,Dill2011}, and we used the relationship~\cite{Ghosh2009}:
\begin{equation}
 \ln\pt{\frac{k_f}{k_u}}=-\gD G/kT = E_{nat}/kT-N\ln z,
\end{equation} 
 where $z$ is the number of rotational degrees of freedom per peptide bond.  To account for the 
difference between the 2D model and real 3D proteins, we calibrated the parameters taken from the 
literature to yield unfolding/folding rates that are meaningful in the context of the other rates 
in our model: folding is much faster than growth and for any of the sequence in our pool 
$k_f/k_u\in (10^2,10^4)$~\cite{Ghosh2010,Dill2011} for 3D proteins.  Because the literature models 
are only mean-field, averaged over sequences, and in order to retain sequence dependence here, we 
set the unfolding rate of all sequences to the average for their lengths, and assigned all the 
sequence dependence to $k_f$.  So, we used: 
significantly.
\begin{equation}
\begin{split}
  k_u &= \exp[12-0.1 \sqrt{N} -E_H(0.5 N + 1.34)],\\
  k_f &= k_u\exp(\gD G)
\end{split}
\end{equation}
The model is not sensitive to varying these parameters over a wide range.  We use $E_h \approx 
1-2$kT, so $k_{unf} \approx 10^2$, which leads to a range of unfolding rates from  a reaction per 
hours and days and range of folding rates from a 
reaction per hours to fractions of a second.

\item The catalytic step is:
\begin{equation}
\mbox{Catalyst}+H+ \underbrace{\cdots HH}_{l-1} 
\xrightarrow{\beta_{cat}}\mbox{Catalyst}+\underbrace{\cdots HHH}.
\end{equation}
 The rate enhancement is $\beta_{cat}=\beta\cdot\exp(E_{h}\cdot n_{c}/kT)$, where hydrophobic 
sticking energy is $e_H$, the number of contacting hydrophobes is $n_c$, which varies in the range 
$3-6$. With the hydrophobic energies of $e_H = 1-2$kT, this gives catalysis rates around hours to 
days per reaction. Because the EPDM supports only binary reactions, we divided the reaction above 
into to steps: interaction of catalyst with a monomer with rate $\beta$ and the reaction of this 
complex with a polymer has the rate $\beta_{cat}$. 
\end{itemize}

For each trajectory, we collected statistics only after the system reached an unchanging steady 
state.  In order 
to explore the stochasticity, we repeated every simulation for 30 times for every experiment. We 
ran all the simulations for $140s$ of internal simulation time, during which $10^6-10^9$ 
individual 
reactions had occurred. We took measurements every $10^{-6}s.$ 
For all the trajectories steady state behavior was reached no longer than $40s$ after the start of 
a simulation. Thus we considered only the last $100s$ (one million recordings) for each 
simulation. 
All the data points we used in the figures are averages over these recordings.

For all the experiments below, we used the following parameters:
\begin{enumerate}
 \item $\beta = 1$
 \item $E_h = 2kT$
 \item $z=1.2$
 
 \item $a=1000$
 \subitem Values of $a\ll 1000\,\,\mbox{or}\,\,a\gg1000$ are problematic, having numbers of 
sequences or populations either too high to calculate or too low to draw conclusions.
 
 \item $h=d=0.1$.
 \subitem When $3d\lessapprox h\leq\beta$, hydrolysis is dominates and without catalysis, there's 
an 
explosion of short sequences. 
\subitem When $3h\lessapprox d\leq\beta$, hydrolysis is unphysically small, so nothing limits the 
growth of longer sequences, even in the absence of catalysis. 
\subitem When $0.05\lessapprox d\approx h \lessapprox 0.5$ the forces of dilution and hydrolysis 
are relatively balanced and populations are neither too small or too large.
 
\end{enumerate}

\subsection*{\textit{In-silico} experiments}\label{sec:experiments}
The simulations were performed on the Laufer Center's computing cluster of CPUs. 
Source files of the models, parameters, initial conditions and random seeds can be obtained at 
\url{https://github.com/gelisa/hp_world_data}.  We performed the following computational 
experiments:

\textbf{Experiment 1: Does our bare polymerization reproduce the Flory 
distribution?}\label{sec:expt1}
We started simulations with a small pool of chains up to 3-mers. To calculate the length 
distributions, 
we calculated for each trajectory the average population of every sequence over time over all 
recordings after 40s, resulting in a million time steps.  Then we summed all the populations of a 
given 
length, obtained total populations for all $n$-mers, $n\in[1,25]$, and then computed every 
population as:
\begin{equation}
 p_n = \frac{\sum\mbox{all n-mers}}{\sum\mbox{total population}}
\end{equation}
giving probability of finding an $n$-mer of a randomly chosen molecule in the system.

The source file of the model and parameters of the simulation are located at 
\url{https://github.com/gelisa/hp_world_data/tree/master/001}

\textbf{Experiment 2. What is the effect on the distribution of just HP folding?}
We started with the same initial population as in Experiment 1. But now we introduce the 
hydrophobic 
energy $e_h= 2kT$. To calculate the resultin length distribution, we computed the average 
population 
of every sequence for each trajectory over time over all the recordings after 40s, resulting in a 
million time steps. The source file of the model and parameters of the 
simulation are located at \url{https://github.com/gelisa/hp_world_data/tree/master/002}

\textbf{Experiment 3. What is the effect on the distribution of both folding and catalysis?}
In addition to folding in this \textit{in-silico} experiment, we also accounted for the pairwise 
contact interactions between two proteins, with the parameters as indicated above. We explored 
ranges of parameters.  We observed significant stability of the length distribution towards change 
of $h$ and $d$: $0.05\lessapprox d\approx h \lessapprox 0.5$.  The 
distributions we observe are quite sensitive to the choice of hydrophobic energy, as expected for 
chemical reactions, since this enters into the exponent of the rate expression. In the generally 
physical range of $e_h= 1-3 kT$, we observe a bending of the Flory distribution, as noted in the 
text. The source file of the model and parameters of the 
simulation are located at \url{https://github.com/gelisa/hp_world_data/tree/master/003}

\bibliography{library}
 \bibliographystyle{achemso}
\end{document}